\documentclass[fleqn,usenatbib]{mnras}
\usepackage{txfonts}
\usepackage{xcolor}
\usepackage[T1]{fontenc}
\usepackage{ae,aecompl}
\usepackage{changepage}
\usepackage{graphicx}	% Including figure files
\usepackage{amssymb}	% Extra maths symbols
\usepackage[
singlelinecheck=false % <-- important
]{caption}
\title[Polarization Observation of Venus]{Polarisation and Brightness Temperature Observations of Venus with the GMRT}

\author[Mohan et al.]{
Nithin Mohan,$^{1}$
Suresh Raju C,$^{1}$\thanks{E-mail: c\_sureshraju@vssc.gov.in}
Govind Swarup$^{2}$
and Divya Oberoi$^{2}$
\\
$^{1}$Space Physics Laboratory, Vikram Sarabhai Space Centre, ISRO, Thiruvananthapuram 695 022, India \\
$^{2}$National Centre for Radio Astrophysics, TIFR, Pune University Campus, Post Bag 3, Pune 411 007, India	\\
}

\date{Accepted XXX. Received YYY; in original form ZZZ}

\pubyear{2018}

\begin{document}
\label{firstpage}
\pagerange{\pageref{firstpage}--\pageref{lastpage}}
\maketitle

\begin{abstract}
Venus was observed at frequencies of 1297.67\,MHz (23\,cm), 607.67\,MHz (49\,cm) and 
233.67\,MHz (1.28\,m) with the Giant Metrewave Radio Telescope (GMRT) during the 
period of 25th July and 6th September 2015 when it was close to its inferior 
conjunction. Values of the brightness temperature ($\textit{T}_\textup{b}$) of Venus from these 
observations were derived as 622$\pm$43\,K, 554$\pm$38\,K for 1297.67 and 607.67\,MHz 
frequencies, respectively, are in agreement with the previous observations. The 
attempt to derive the $\textit{T}_\textup{b}$ at 233.67\,MHz affirms an upper limit of 321\,K which is 
significantly lower than the previously reported upper limit of 426\,K at the 
same frequency. We also present the dielectric constant ($\varepsilon$) values of the Venus 
surface estimated using the degree of polarisation maps of Venus, derived from 
the GMRT polarisation observations and theoretical calculations. The $\varepsilon$ of the 
Venus surface was estimated to be $\sim$4.5 at both the 607.67\,MHz and 1297.67\,MHz, 
close to the reported values of $\varepsilon$ of 4 to 4.5 from the radar-based observations 
including the Magellan observations at 2.38\,GHz (12.6\,cm).

\end{abstract}

% Select between one and six entries from the list of approved keywords.
% Don't make up new ones.

%%%%%%%%%%%%%%%%%%%%%%%%%%%%%%%%%%%%%%%%%%%%%%%%%%

%%%%%%%%%%%%%%%%% BODY OF PAPER %%%%%%%%%%%%%%%%%%

\section{Introduction}

Observations of the planetary bodies at microwave and radio wavelengths 
(millimetre (mm), centimetre (cm) and decimetre (dcm)) have made significant 
contributions towards understanding the planetary atmospheres and their surfaces 
\citep{dePater1990, dePater1991}. The albedo of the planetary surface determines 
the fraction of the solar radiation reflected from the surface and the rest is 
conducted into the subsurface as heat. Depending upon the surface properties 
(dielectric constant and surface roughness) and the local thermal equilibrium, 
the energy is then re-radiated in the form of thermal energy. The net balance 
between the former, the latter and the incident solar radiation determines the 
thermal evolution of the planetary subsurface \citep{dePater1990, dePater1991}. 
However, the planetary thermal emissions do not strictly follow a blackbody 
curve due to the frequency dependence of the surface emissivity and the 
temperature gradients in the atmosphere or subsurface \citep{Kellermann1966}. Venus 
has a thick atmosphere dominated by CO$_2$ ($\sim$96\%).  Due to the resulting 
Greenhouse effect, the surface temperature of Venus is $\sim$735\,K.  The 
microwave/radio observations at cm/dcm wavelengths have the potential to probe 
deeper into the planetary atmosphere and its surface. The thermal radiation 
from the planet has contributions from its atmosphere and the surface beneath 
it. The atmosphere is transparent enough at longer wavelengths that it permits 
the probing of planetary subsurface \citep{Muhleman1973, Warnock1972, Butler2001}.
Also, at the long wavelengths, the surface 
reflectivity of the Venus is nearly a constant and is fairly low ($\sim$10-15\%) 
\citep{Goldstein1965}, markedly reducing the component of atmospheric emission 
reflected from the surface.

The presence of thick atmosphere and acid clouds prevent the surface probing 
apart from the IR windows between 0.9 to 2.5\,$\mu$m \citep{Allen1984, Meadows1996}.
Spectroscopic techniques have been used to study the 
Venusian upper atmosphere and its constituents as well as the presence of H$_2$SO$_4$ 
droplets of $\sim$1\,$\mu$m size in the Venusian clouds \citep{Hansen1974, Kawabata1975}.
The VIRTIS probe onboard Venus Express mission had 
dedicated channels of 0.9, 1.1, 1.18\,$\mu$m to study variations of the surface skin 
temperature of Venus \citep{Drossart2004} and were used to identify hotspots 
on the surface of Venus, which form the direct evidence of recent volcanic 
activities \citep{Smrekar2010}.

\citet{Mayer1958} pioneered the microwave probing of the Venus disk at 3.15\,cm 
wavelength and reported an unexpectedly warm temperature of 560$\pm$73\,K, 
following this, several observations were carried out of the planet over a wide 
range of wavelengths in the mm, cm and meter regime. The initial observations of 
Venus were in visible spectra and had the popular assumption about the planet 
being similar to Earth. The mm/sub-mm wavelength observations are ideal for 
probing the planetary middle and upper troposphere \citep{Devaraj2011}. 
These have been used to study the Sulfur bearing molecules which affect the 
thermodynamics of the Venusian atmosphere \citep{Butler2001}.
At shorter wavelength regime (mm-cm), the emission is largely 
dominated by the colder region of the Venus atmosphere \citep{dePater1991}. At the 
longer cm-dcm wavelengths, the emission has the signature of the surface and 
subsurface thermo-physical and electrical properties of the regolith. The values 
of the brightness temperature, $\textit{T}_\textup{b}$, increases from $\sim$505\,K 
at 1.3\,cm to 680\,K at 
6.2\,cm and then decreases at longer wavelengths \citep{Butler2001}. The recent 
radiometric observation at higher dcm/m wavelength observations using the GMRT 
reported $\textit{T}_\textup{b}$ values of 526$\pm$22\,K, 409$\pm$33\,K, and 
$<$426\,K at 49\,cm, 90\,cm and 
1.23\,m, respectively, which indicate further decrease in the observed $\textit{T}_\textup{b}$  
\citep{Mohan2017}.

Neglecting atmospheric contributions, the radiometric observations of 
Venus measure its blackbody equivalent 
temperature or brightness temperature, which is a function of surface dielectric 
properties, temperature and roughness. While measurements of the total intensity 
(Stokes parameter, I) are related to the $\textit{T}_\textup{b}$ of a planetary disk, the 
polarimetric observations are necessary to derive all the four Stokes parameters 
(I, Q, U, V) for studying surface properties like the dielectric permittivity 
and roughness. The emission from the surface of a planetary body is polarised 
due to the difference in the two orthogonal components of the electric field 
vector (perpendicular and parallel to the plane of incidence) at the air-surface 
interface \citep{Troitskii1954, Clark1965}. The difference between these 
two orthogonal components at the larger angle of incidence is manifested as an 
enhancement in the degree of polarisation (DOP) towards the limb. Also, the 
emission becomes depolarised with increasing surface roughness \citep{Pollack1965}.
\citet{Troitskii1954, Davies1966} used the polarisation 
observations of the lunar surface to derive its dielectric constant ($\varepsilon$) and the 
surface roughness, which was followed by \citet{Heiles1963} for deriving the 
surface dielectric constant of Venus. Several radio and radar observations were 
conducted to derive the dielectric properties of the Venus surface. Assuming the 
Venusian surface to be significantly smooth at cm/dcm wavelengths \citep{Carpenter1964},
\citet{Clark1965} reported the $\varepsilon$ of Venus $\sim$2.5 at 10.6\,cm 
wavelength. Later, \citet{Muhleman1979} estimated the $\varepsilon$ as 4.1 based on: (i) 
value of the radio brightness of Venus measured with the Owens Valley 
interferometer, (ii) its radar reflectivity using the Arecibo radar and (iii) 
Mariner 5 and 10 occultation observations. \citet{Chapman1986} measured the degree of 
polarisation across Venus from observations made at 21\,cm using the Very Large 
Array (VLA) and obtained a value of 2.15. \citet{Pettengill1988} found that 
the lowland regions on Venus are characterized by $\varepsilon$ = 5.0$\pm$0.9 from the 
Pioneer-Venus radar altimetry experiment. Further, \citet{Pettengill1992} used 
the Magellan radiometer observations at 12.6\,cm wavelength and determined the 
global mean value of its emissivity as 0.845 that corresponds to a dielectric 
permittivity of between 4.0 and 4.5 depending on its surface roughness.

This paper describes the observations of Venus made close to its inferior 
conjunction in 2015 with the Giant Metrewave Radio Telescope (GMRT) with two 
objectives: (a) to determine values of its brightness temperature at 1297.67\,MHz 
($\sim$23\,cm), 607.67\,MHz ($\sim$49\,cm) and $\sim$ 233.67\,MHz ($\sim$1.28\,m)
and (b) to estimate 
value of the dielectric constant of its surface from polarisation observations 
at 1297.67\,MHz and 607.67\,MHz. Section 2 describes the GMRT observations, 
followed by a description of the data reduction and analysis procedure in 
Section 3. The results are described in Section 4. Sections 5 and 6 present the 
discussion and conclusions, respectively.

\section{Observation} 

Observations of Venus were carried out on four days, near its inferior 
conjunction in 2015, with the GMRT at frequencies centred at 233.67\,MHz, 607.67\,MHz
and 1297.67\,MHz (Table \ref{tab:Table-1}). The GMRT is an aperture synthesis
radio telescope 
consisting of 30 fully steerable parabolic dish antennas, each with a diameter 
of 45\,m. Fourteen dishes are located in a central array of $\sim$1\,km size and the 
other sixteen along three arms of a nearly Y-shaped array $\sim$25\,km in extent 
\citep{Swarup1991}. Dual orthogonal dipole feeds operating at 606, 325, 235 
and 150\,MHz measure linear polarisations which are converted to the left and 
right-handed circularly polarised signals using a quadrature hybrid. The horn 
type antennas operating in the 1000-1450\,MHz frequency band directly measure 
orthogonal circular polarisations, and a quadrature hybrid is used to convert 
them to X and Y linear polarised signals. The GMRT correlator can provide all 
four cross-products for both linearly and circularly polarised signals and 
allows determination of all Stokes parameters, I, Q, U and V \citep{Chengalur2013}. 

Full polarimetric data was recorded for the two higher frequencies, and only 
total intensity (Stokes I) data was recorded for 233.67\,MHz  as the polarised 
emission at this frequency is insignificant compared to the noise contributed by 
the sky background and the electronics system. The observation procedure 
described by \citet{Mohan2017} was adopted for the current observations. 
Instead of tracking Venus, whose right ascension (RA) and declination (Dec) 
varied during the observations, the GMRT antennas tracked the mean position of 
its RA and Dec. For weak non-sidereal sources, this is the preferred strategy as 
it allows one to use the background celestial sources present in the field of 
view for self-calibration \citep{Cornwell1999}. Self-calibration 
allows one to take into account the time-varying complex antenna gains (phase 
and amplitude), resulting from variations of  the gains of the GMRT electronics 
and those due to the ionosphere, and leads to a much higher imaging dynamic 
range and fidelity. This is especially helpful at low radio frequencies, where 
the Galactic synchrotron background as well as the bulk of cosmic extragalactic 
source grow increasingly brighter due to their negative spectral index. The 
ephemeris details of Venus during the observations are given in Table \ref{tab:Table-2}. These 
observations were timed to be close to the inferior conjunction of Venus, when 
the observed flux density of Venus would be the largest and so would be the 
apparent size of the disc of the planet. This led to an improved SNR and a 
larger number of synthesized beams across the planet, and thus enabled the most 
detailed and sensitive observations possible with the GMRT. Table \ref{tab:Table-3} gives 
details of the calibrator sources used for the GMRT observations. The compact 
radio sources 0943-083, 1021+219, 0842+185 and 0842+185 were chosen for 
calibrating phases. For polarisation calibration, a single scan of 3C138 was 
observed near the middle of observations for $\sim$10 min and also a single scan of 
3C286 was made at the end of observations of Venus. Values of the fractional 
polarisation and polarisation angles for 3C138 and 3C286 used were given by 
\citet{Perley2013} and are listed in Table \ref{tab:Table-4}.

\begin{table*}
  
  %\begin{center}
  \caption{GMRT campaign for observations of Venus 2015}
  \label{tab:Table-1} 
  \centering\small
  \begin{tabular}{cccccc}
\\
\hline
 
  Central  	 & Date  	& Bandwidth     & No. of   & Observing   & Stokes	\\
  Frequency	 &	 	&               & working  & time on	 &		\\
  (MHz)	   	 &	 	& (MHz)  	& antennas & Venus (mins)&		\\
  \hline 
  
    607.67	 & July 25 	& 33.33 	&  28	   & 5.5	  & I, Q, U, V	\\
    233.67	 & July	26	& 16.67 	&  27	   & 6.5	  & I		\\
    607.67	 & August 29	& 33.33 	&  28	   & 6.0	  & I, Q, U, V	\\
    1297.67 	 & September 06	& 33.33 	&  28	   & 6.5	  & I, Q, U, V	\\

  \hline	      
  %\bottomrule
  \end{tabular}	      
\end{table*}

\begin{table*}
\caption{Ephemeris data for GMRT campaign for observations of Venus 2015}
\label{tab:Table-2}
\begin{center}
\centering\small
\begin{tabular}{cccccccc}
  \hline
  
   Central	& Date		&  Venus Angular  	  & Right Ascension 	& Declination		& Sub Earth	 & Sub Earth 		& Phase	\\
   Frequency	& 		&  Diameter\,($\theta$)   & (mid-observation)   & (mid-observation)	& Latitude 	 & Longitude 		&	\\
   (MHz)	& 		&  (arcsec)		  & (h:min:s)		& (deg.arcmin.arcsec)	& 		 & (deg East)		&	\\
 \hline
  607.67	&  July	25	&  47.43	 	  &	10:05:12	& +07.47.39		& 4.34	   	 &	307.6		& 0.13	\\
  233.67	&  July 26	&  48.16	 	  &	10:04:49	& +07.35.31		& 4.56	   	 &	309.06		& 0.12	\\
  607.67	&  August 29	&  53.33	 	  &	09:02:25	& +08.37.39		& 8.23	   	 &	344.16		& 0.068	\\
  1297.67	&  September 06	&  48.02	 	  &	08:58:21	& +09.51.16		& 7.31	   	 &	354.65		& 0.14  \\
		
  \hline
%\bottomrule
  \end{tabular}

  \end{center}
\end{table*}

\begin{table*}
\caption{Details of calibrators for GMRT campaign for observation of Venus 2015}
\label{tab:Table-3}
\begin{center}
\centering\small
\begin{tabular}{ccccc}
 \hline
Date		& Frequency	& Flux 		& polarization  & Phase      \\
		& (MHz)		& calibrator	& calibrator	& calibrator \\
\hline
July 25		& 607.67	& 3C147, 3C286	& 3C286, 3C138	& 0943-083   \\
July 26		& 233.67	& 3C147, 3C286	& 		& 1021+219   \\
August 29	& 607.67	& 3C147, 3C286	& 3C286, 3C138	& 0842+185   \\
September 06	& 1297.67	& 3C147, 3C286	& 3C286, 3C138	& 0842+185   \\
\hline
  
%\bottomrule
  \end{tabular}

  \end{center}
\end{table*}

\begin{table*}
\caption{Detailes of polarization properties of Calibrator sources for observation of Venus 2015}
\label{tab:Table-4}
\begin{center}
\centering\small
\begin{tabular}{ccccccc}
  \hline
Freq 		&  3C138   		&      		& 3C286   	  &		\\
(MHz)		&  \% Pol  		& Pol ang   	& \% pol  	  & Pol ang 	\\
  \hline
607.67		&  3.9	   		& -26		& 7.6	  	  & 33		\\
1050.0		&  5.6	   		& -14		& 8.6		  & 33       	\\
1297.67		&  6.7			& -12	    	& 9.1	  	  & 33		\\
1450.0		&  7.5	   		& -11		& 9.5		  & 33        	\\

  \hline                                                                
%\bottomrule                                                                                             
  \end{tabular}

  \end{center}
\end{table*}

\section{Data Analyses}
Except for the polarisation calibration, the procedures involved and followed in 
the data analyses are detailed in \citet{Mohan2017}. Brief discussions of 
these procedures are given in this Section for completeness. GMRT data were 
analysed using the Common Astronomical Software Application (CASA) package   
\citep{McMullin2007}. An integration time of 2 second was chosen to 
facilitate the removal of the fast varying radio frequency interference and 
then averaged to 16 seconds. A spectral resolution of 125\,kHz was used for the 
observations. The polarisation observations at low frequencies are sensitive to  
variations of the ionospheric total electron content (TEC) over the GMRT line 
of sight, which can be  determined from the GNSS and GPS measurements 
(\url{ftp://cddis.gsfc.nasa.gov/gnss/products/ionex/}). Using CASA recipes, the global 
TEC variation map over the GMRT sky for these observation days were generated 
and then applied to the measurement set using the task GENCAL. \citet{Perley2013}
flux density scale was applied for observations made at 607.67\,MHz and 
1297.67\,MHz, whereas \citet{Scaife2012} flux density scale was used for 233.67\,MHz
observations. The initial frequency and time-dependent complex gain 
solutions were obtained using the tasks BANDPASS and GAINCAL. Polarisation 
calibration was done prior to scaling the flux of the compact calibrator sources 
(task FLUXSCALE). Standard steps recommended for routine polarimetric continuum 
analysis were followed (e.g. as described in 
\url{https://casaguides.nrao.edu/index.php/VLA_Continuum_Tutorial_3C391-CASA5.0.0} for 
the analysis of VLA P-band continuum data), and are briefly described below.

\textbf{Polarisation calibration} was performed for removing instrumental contribution to 
the observed polarisation. This involved solving for the unknown instrumental 
polarisation as well as calibrating the absolute polarisation position angle.  
After the task SETJY the model fluxes of the Stokes parameters Q and U of the 
polarisation calibrators were derived from the relation:\\
\\
$I_0 = Total\,\,Intensity $     	\\
$Q_0 = P_0\,cos\,\,(2\chi \pi/180)$ 	\\
$U_0 = P_0\,sin\,\,(2\chi \pi/180)$	\\

where $P_0$, the fractional polarisation and $\chi$ = polarisation position angle for 
3C138 and 3C286 (Table \ref{tab:Table-4}). Using these parameters, the polarisation models were 
generated for 3C138 and 3C286. The cross-hand delay calibration using the strong 
unpolarised source 3C147, to account for the delay difference between the two 
orthogonal polarisations of the reference antenna was carried out. Any antenna 
has some polarisation leakage associated with it. This, along with the 
imperfections in the electronics chain, gives rise to a spurious polarisation, 
which makes even an unpolarised source appear somewhat polarised. This frequency 
dependent instrumental leakage term (D-term) was determined using the source 
3C147. In order to obtain accurate polarisation position angle, the two 
orthogonal phases  need to be calibrated. For this, the phase calibrator having 
significant parallactic angle coverage was utilized due to the fact that the 
point source is not 100\% unpolarised. Afterwards, the flux of the phase 
calibrator sources was determined using the task FLUXSCALE. The parallactic 
angle corrections for all the data were done while applying the calibration 
solutions to the measurement set, using the PARANG parameter in APPLYCAL.

\textbf{Imaging:} The background celestial radio sources in the field of view of the GMRT 
antennas were used for the self-calibration as described by \citet{Mohan2017} 
in the imaging procedure Section. The procedure was continued until the gain 
phases varied within $\pm$5 degrees. Next, the CLEAN components were subtracted from 
the maps using the task UVSUB. One-minute snapshots of Venus in all the four 
Stokes parameters were generated after compensating for the non-sidereal motion 
of Venus. The co-added maps of Venus were then deconvolved using the point 
spread function corresponding to the entire observation, followed by a 
convolution using the corresponding restoring beam. The uncertainty in the 
estimated flux density is contributed by - pointing errors, temporal variation 
of the gain of the system and the errors due to uncertainties in flux values of 
the calibrators due to flux density scale used. The detailed error estimation is 
presented in \citet{Mohan2017}. All of the above factors together can lead to 
systematic error of about 7\% in the measured flux density of Venus. 
    
Besides the above corrections, the Galactic background temperature ($\textit{T}_\textup{gal}$), 
corresponding to the location of Venus on the day of the observation was 
estimated from the 408\,MHz map by \citet{Haslam1982}. A spectral index of -2.6$\pm$0.15 
\citep{Reich1988} was used to scale the $\textit{T}_\textup{gal}$ values corresponding to the GMRT 
frequencies. The values of $\textit{T}_\textup{gal}$ were determined as 6$\pm$1 K, 5$\pm$1 K, and 64$\pm$8\,K for 
July 25 (607.67\,MHz), August 29 (607.67\,MHz) and July 26 (233.67\,MHz) 
observations, respectively. For the 607.67\,MHz and 1297.67\,MHz observations, the 
background temperature is a sum of the galactic background and cosmic microwave 
background temperature, which are $\sim$6+2.7 $\sim$9K, $\sim$5+2.7 $\sim$8K, $\sim$1+2.7 $\sim$4K for July 
25 (607.67\,MHz), August 29 (607.67\,MHz) and September 06 (1297.67\,MHz), 
respectively, are also accounted for. 

The consistency of the images was then evaluated by grouping the 1-minute 
snapshots into two time bins and independently coadding. It was found that, for 
25th July and 29th August (607.67\,MHz) the observed the flux density was 
consistent within 2\% while for 1297.67\,MHz and 233.67\,MHz the flux densities 
were consistent within a factor of $\sim$3\% and $\sim$5\%, respectively.

\section{Results}
\subsection{Brightness temperature of Venus}
 
The GMRT observations of Venus were made in 2015, with primary 
objective of observing the planetary thermal emission for a longer period during 
the inferior conjunction of Venus, when it is closest to Earth. During this 
observation period, the planet had an angular diameter of $\sim$50$^{\prime\prime}$ which is almost 
twice that during the 2004 observations, and the observation duration was also 
almost double that of the 2004 observations. All these factors have 
significantly improved signal to noise ratio while extending the GMRT 
observation towards a higher frequency, at 1297.67\,MHz. Table-\ref{tab:Table-5}
presents the results of the flux density and $\textit{T}_\textup{b}$ measurements of the Venus. 
Columns 3 and 4 give the size and the position angle of the synthesized beam respectively. 
Column 5 gives the observed background \textit{rms} values in the map of Venus. There is 
an improvement of $\sim$5 times in the signal to noise ratio in the maps derived at 
607.67\,MHz from the 2015 observations, compared to that from the 2004 
observations, owing to the closer proximity of the planet. Columns 6 and 7 give 
the measured flux density of Venus and the estimated error of its flux density. 
Column 8 gives values of $\textit{T}_\textup{b}$, that were derived using the Rayleigh-Jeans 
relation, measured flux density and solid angle subtended by Venus. Estimated 
values of $\textit{T}_\textup{gal}$ are given in Column 9. Column 10 gives the corrected values of 
Venus $\textit{T}_\textup{b}$ by adding $\textit{T}_\textup{gal}$ values and microwave background temperature value for 
the 233.67\,MHz, 607.67\,MHz and 1297.67\,MHz observations. In Column 11 are given 
estimated values of the $\textit{T}_\textup{b}$ in bold and also their estimated errors as derived by 
adding the quadrature sum of the random errors as determined from column 7 and 
the systematic error of 7\%. 

The total intensity maps of the Venus were generated for the respective days of 
observations. The observed $\textit{T}_\textup{b}$ variation ranges from $\sim$100\,K near the limb, 
peaking at $\sim$650\,K around the centre of the disk for the 25 July and $\sim$100\,K to 
$\sim$700\,K for the 29 August 607.67\,MHz observations. In case of 1297.67\,MHz, the $\textit{T}_\textup{b}$ 
ranges from $\sim$200\,K near the limb to $>$750\,K near the centre of the disk, 
indicating a brighter Venus at 1297.67\,MHz frequency. As described in Table \ref{tab:Table-5}, 
the mean values of $\textit{T}_\textup{b}$ are found as 554$\pm$38\,K at 607.67\,MHz and 622$\pm$43\,K at 
1297.67\,MHz, respectively.  

In the case of the 233.67\,MHz observations made on 26 July 2015, a strong 3C 
source of $\sim$10 Jy was present near the half maximum position of the GMRT primary 
beam limiting the dynamic range of the deconvolved image, whence Venus was 
buried in the map noise. Hence, an upper limit of 3$\sigma$ was chosen for the expected 
flux density of Venus at 233.67\,MHz as shown in row 3, column 6 in Table \ref{tab:Table-5}.  

The $\textit{T}_\textup{b}$ values of Venus obtained from the 2015 observations are plotted 
in Fig. 1 along with the $\textit{T}_\textup{b}$ observations made with the VLA by \citet{Butler2001}
and those using the GMRT by \citet{Mohan2017}. These observations show 
that $\textit{T}_\textup{b}$ peaks at $\sim$680\,K at a wavelength of 6\,cm and then decreases steadily to 
622\,K at 23\,cm, 554\,K at 49\,cm, 409\,K at 91\,cm and to $<$321\,K at 128\,cm. The $\textit{T}_\textup{b}$ 
values obtained by GMRT at 1297.67\,MHz are consistent with the VLA observations 
made by \citet{Butler2001}. Also, the $\textit{T}_\textup{b}$ at 607.67\,MHz agrees well with that 
reported by \citet{Mohan2017} based on GMRT observations in 2004. The result 
obtained from the current observations at 233.67\,MHz establish that the $\textit{T}_\textup{b}$ 
continues to decrease even at wavelengths beyond one metre.

\begin{table*}
  \begin{center}
  \caption{Summary of $\mathit{T}_\textup{b}$ measurements (2015)}
  \label{tab:Table-5}
    \centering\small
%  \begin{adjustwidth}{-0.8cm}{}
    \begin{tabular}{cccccccccccc} 
\\
\hline
  Frequency	& Date   	  & Beam Size 				    			 & Beam 	   	& Map		& Flux           & $\Delta$ Flux    &  $\mathit{T}_\textup{b}$(K)	& $\mathit{T}_\textup{gal}$(K)	 & $\mathit{T}_\textup{b,cor}$ (K)  	& mean 				\\
  (MHz)		& (2015)   	  &  (arcsec)				    			 & Position 		& $\textit{\textit{rms}}$& (mJy)	 	 & (mJy)	    &					& 				 &					& $\mathit{T}_\textup{b,cor}$	\\	
		&	  	  &	   				    			 & Angle\,(degree) 	& (mJy)		&		 &	 	    &					&				 &	  	 	     		&(K)				\\
  \hline
  233.67	& Jul 26 	  & 12.66$^{\prime\prime}$ $\times$ 10.8$^{\prime\prime}$	 & 55.1 		& 1.52		& 18.18	    	 & 6.27	            & $\downarrow$ 249			& 64$\pm$8 			 & \textbf{$\downarrow$321}		& \textbf{$\downarrow$321}	\\
  607.67	& Jul 25 	  & 5.67$^{\prime\prime}$ $\times$ 5.03$^{\prime\prime}$ 	 & 62.6			& 0.16		& 251.087	 & 0.75	    	    & 533.06		 		& 6$\pm$1 	 		 & 543 					& \textbf{554$\pm$38}		\\
  607.67	& Aug 29	  & 6.14$^{\prime\prime}$ $\times$ 4.28$^{\prime\prime}$ 	 & 46.8			& 0.21		& 330.36	 & 1.10	            & 557.23		 		& 5$\pm$1 			 & 566      				& 				\\ 
  1297.67	& Sep 06	  & 2.98$^{\prime\prime}$ $\times$ 2.22$^{\prime\prime}$ 	 & 61.9			& 0.22		& 1354.00	 & 2.05	            & 618.13		 		& 4 			 	 & 622      				& \textbf{622$\pm$43}		\\

 \hline

   \end{tabular}
%  \end{adjustwidth}
\end{center}

\end{table*}

\begin{figure}
  \centering
  \includegraphics[width=0.5\textwidth]{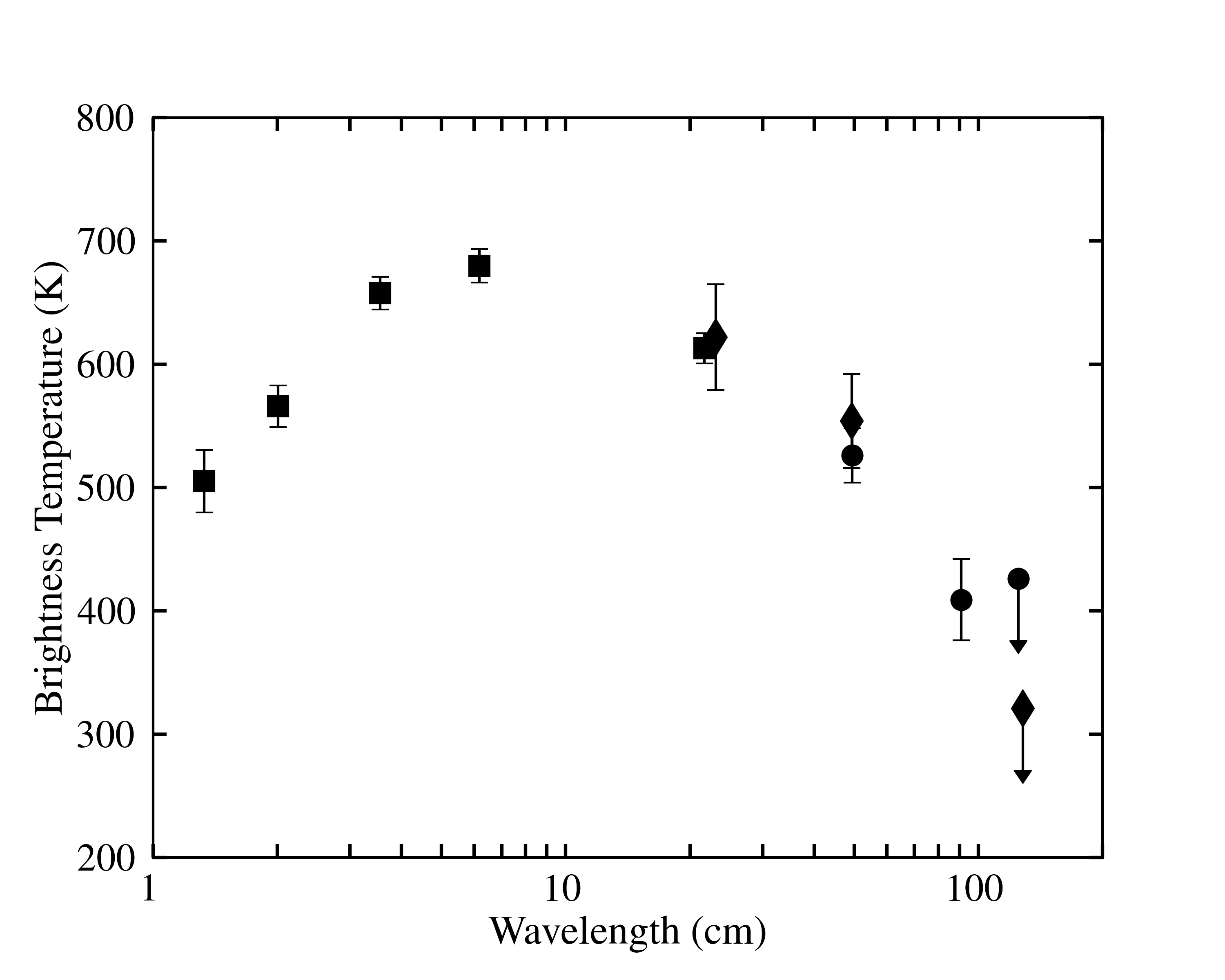}
    \captionsetup{font=footnotesize}
  \caption{
	   Summary of T$_b$ measurements on Venus in cm-dcm wavelengths. The solid square ($\blacksquare$) represents the 
	   observation by \citet{Butler2001} using the VLA. The solid circles ($\bullet$) represent the observation by 
	   \citet{Mohan2017} using the GMRT and the current observations are illustrated by the solid diamond ($\blacklozenge$)
	   points.
	   }
\label{fig:Fig4}
\end{figure}

\subsection{Degree of polarization measurement of Venus}

The degree of linear polarisation (or fractional polarisation) of any planetary 
surface medium represents the fraction of the linearly polarised component of 
the total intensity of radiation. The properties of thermal radiation (including 
polarisation) are dependent on the surface roughness, surface dielectric 
permittivity and its discontinuity in the horizontal and vertical directions 
within the penetration depth of the wave. Therefore, the degree of polarisation 
(DOP) measurements are carried out to study the dielectric properties and 
surface roughness parameters. The fraction of DOP can be obtained from the 
observations of stokes I, Q, and U parameters as:   

\begin{equation}
\label{eq:Eq1}
 m = \frac{\sqrt{Q^2+U^2}}{I}
\end{equation}
where the numerator corresponding to the linear polarization intensity (P = 
$\sqrt{Q^2+U^2}$) is the total intensity of the linear polarized component of 
the radiation. The Stokes parameters measured at 607.67\,MHz and 1297.67\,MHz
are used to derive the DOP of Venus during the computation.

Figures 2 and 3 show the variations of the DOP across the disk of Venus from the 
observations made at 607.67\,MHz on 25 July 2015 and at 1297.67\,MHz on 06 
September 2015. Division of the polarisation intensity by Stokes parameter I 
lead to exceptionally high values of the DOP outside the disk of Venus and has 
no physical significance. This was countered by masking the degree of 
polarisation outside the disk by setting a threshold of 3$\sigma$ in the total 
intensity maps, where $\sigma$ is the \textit{rms} of intensity map. The DOP increases steadily 
from the centre of the disk towards its limb. At 607.67\,MHz the mean value of 
DOP increased from $\sim$5\% to $\sim$22\%, and that at 1297.67\,MHz from $\sim$5\% at the centre 
to $\sim$32\% towards the limb (refer Section 4.3). The DOP variations, with the 
contour lines overlaid to quantify the DOP are shown in these figures. Under the 
assumptions of a smooth planetary surface with respect to the probing wavelength 
and the magnetic permeability $\mu$ = unity, the dielectric permittivity, $\varepsilon$, of the 
Venus surface can be estimated from the measured DOP values and is described in 
the next Section.
\begin{figure} 
  \centering
  \includegraphics[width=0.45\textwidth]{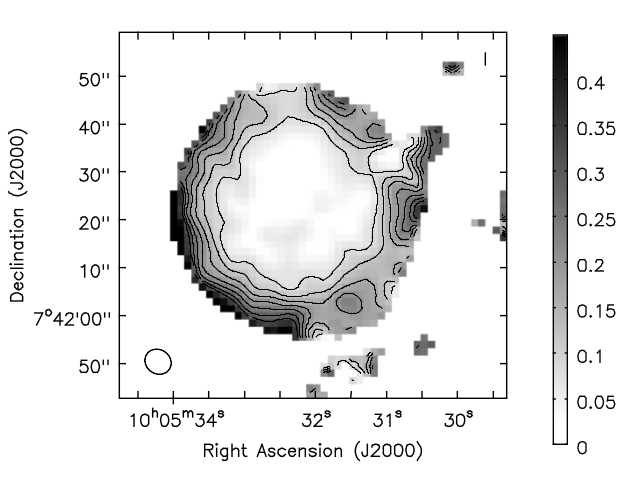}
    \captionsetup{font=footnotesize}
  \caption{
	   Contour plot of degree of polarization across Venus at 607.67\,MHz made on 25 July 2015.
	   Contour Values: 8, 12, 16, 20, 24 and 28 percent
	   }
\label{fig:Fig5}
\end{figure}
\begin{figure}
  \centering
  \includegraphics[width=0.45\textwidth]{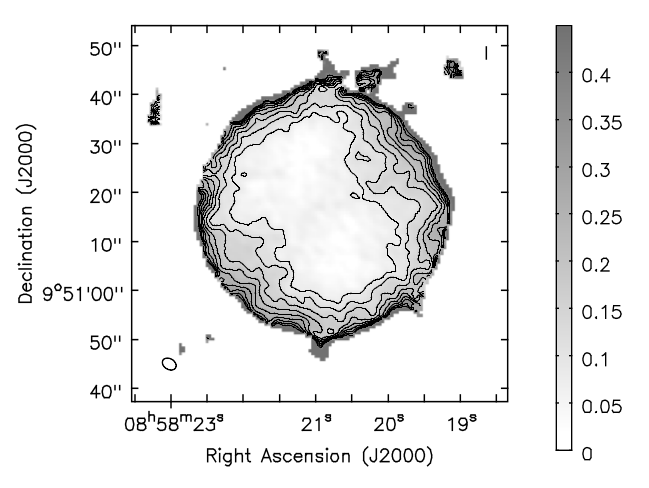}
    \captionsetup{font=footnotesize}
  \caption{
	   Contour plot of degree of polarization across Venus at 1297.67\,MHz on 06 September 2015.
	   Contour values are 8, 12, 16, 20, 24, 28, 32, 36 and 40 percent 
	   }
\label{fig:Fig6}
\end{figure}
\subsection{Estimation of the dielectric constant of Venus surface}

For reflection/refraction from a plane dielectric interface,
the theoretical values of the DOP are related to perpendicular and parallel 
components of the surface reflectivity \citep{Davies1966, Born1980}:
\begin{equation}
\label{eq:EqDOP}
m = \frac{\frac{1}{2}(R_\perp-R_\parallel)}{1-\frac {1}{2}(R_\perp+R_\parallel)}
\end{equation}
where R$_{\perp}$ and R$_{\parallel}$ are, 
respectively, the perpendicular and 
parallel component of power reflection coefficients or the reflectivity of 
the electromagnetic radiation at the plane of incidence and are given 
by the square of the Fresnel's equation. 
\begin{equation}
\label{eq:EqRperp}
R_\perp = \left | \frac{cos\theta_i-\sqrt{(\varepsilon-sin^2\theta_i)}}{cos\theta_i+\sqrt{(\varepsilon-sin^2\theta_i)}} \right |^2 
\end{equation}
\begin{equation}
\label{eq:EqRpar}
R_\parallel = \left | \frac{\varepsilon cos\theta_i-\sqrt{(\varepsilon-sin^2\theta_i)}}{\varepsilon cos\theta_i+\sqrt{(\varepsilon-sin^2\theta_i)}} \right |^2
\end{equation}
where the angle of incidence  $\theta_{i}$ = sin$^{-1}(r/\textit{R}_\textup{disk})$, where `r' 
is the radial distance 
from the centre in arcsec and $\textit{R}_\textup{disk}$ is the angular radius of the disk of Venus in 
arcsec.

The pixel sizes of the images were chosen as 0.6 arcsec and 1.5 arcsec, 
respectively, for the 1297.67 and 607.67\,MHz observations. In the observed DOP 
maps shown in Figs. 2 and 3, we first selected concentric annular regions, each 
of width of 1 arcsec at 1297.67\,MHz and 2 arcsec at 607.67\,MHz observations, 
starting from the centre all the way to the limb of Venus. The angle of 
incidence corresponding to the mean radii of these concentric annular sections 
varied from 0 to a maximum of $\sim$67 degrees from the centre towards the limb at 
607.67\,MHz and that from 0 to a maximum of $\sim$74 degrees at 1297.67\,MHz. The mean 
and the standard deviations of the DOP for each of the annular sections were 
then determined.

It is assumed that each pixel in the annular sections is statistically 
independent. However, since the FWHM of the GMRT beam = $\sim$6 arcsec for 607.67\,MHz 
and $\sim$3 arcsec for 1297.67\,MHz, the beam can occupy $\sim$20 pixels and $\sim$12 pixels 
for 1297.67 and 607.67\,MHz, respectively, on the Venusian disk. But, due to the overlap of the beams on 
the adjacent annuluses the number of independent pixels of DOP occupying the 
beam were determined to be approximately 1/3$^{rd}$ and 1/4$^{th}$ of the total number 
pixels in each annular region, respectively, at 1297.67 and 607.67\,MHz. Later, 
the standard deviation was divided by the revised value of the number of 
independent pixels, \textit{N} and the standard error was found out as Standard 
Deviation/$\sqrt{N}$. The variation of the DOP with increasing radial distance from the 
centre of Venus disk was examined by plotting the mean value of DOP in each 
annular region. In Figs. 4 and 5, using filled circles, are shown the mean 
values of the DOP and the vertical bars are the standard error derived as 
a function of the radial distance from the centre for observations made at 
607.67\,MHz on July 25 and at 1297.67\,MHz on September 6, respectively. The key 
point is that the error bars reflect only the random component of the 
uncertainties in the Figs. 4 and 5, while the departure of the degree of 
polarisation from zero close to the centre of the disk arises because of 
systematics.

These figures also include the theoretical values of the DOP that were derived 
by using the Equations 3, 4 and 5 and then convolved with a Gaussian having FWHM 
corresponding to the synthesized beam of GMRT. These are shown by solid lines 
for four different values of the $\varepsilon$ ranging from 3.5 to 5.5, with increments of 
0.5. The statistical technique of Chi-square minimization is used to identify 
the $\varepsilon$ corresponding to the best fit from the various values ranging from 2.5 to 
6.0 and the value of $\varepsilon$ corresponding to the best fit was obtained as 4.5 at 
607.67 as well at 1297.67\,MHz.

\begin{figure}
  \centering
  \includegraphics[width=0.45\textwidth]{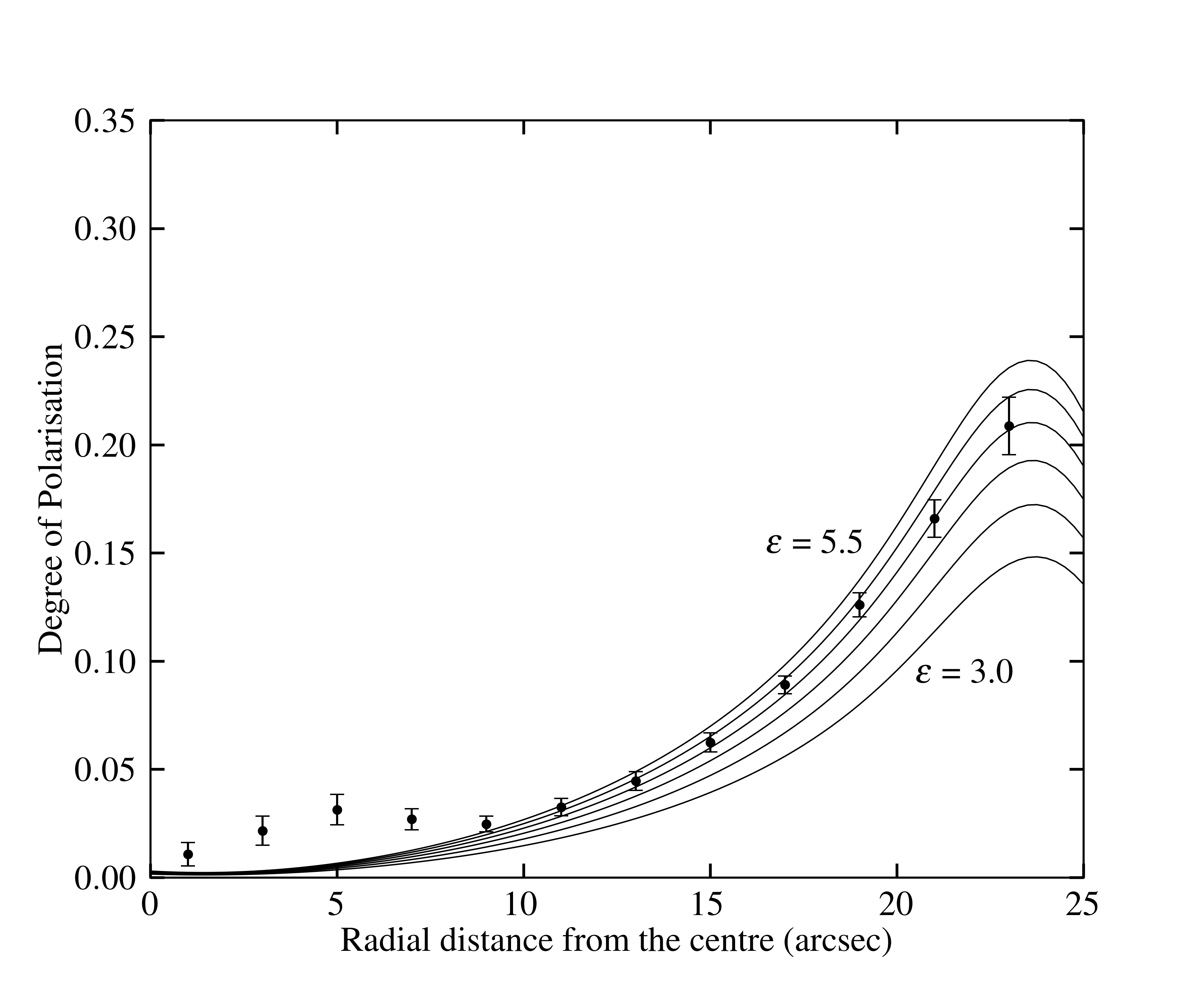}
    \captionsetup{font=footnotesize}
  \caption{
	   Variation of the observed degree of polarization with respect to the radial distance from the centre at 607.67\,MHz as 
	   observed on 25 July 2015 (solid dots). Solid lines show the theoretical curves convolved with Gaussian 
	   corresponding to the GMRT synthesized beam plotted for $\varepsilon$ = 3.0, 3.5, 4.0, 4.5, 5.0 and 5.5
 	   }
\label{fig:Fig7}
\end{figure}

\begin{figure} 
  \centering
  \includegraphics[width=0.45\textwidth]{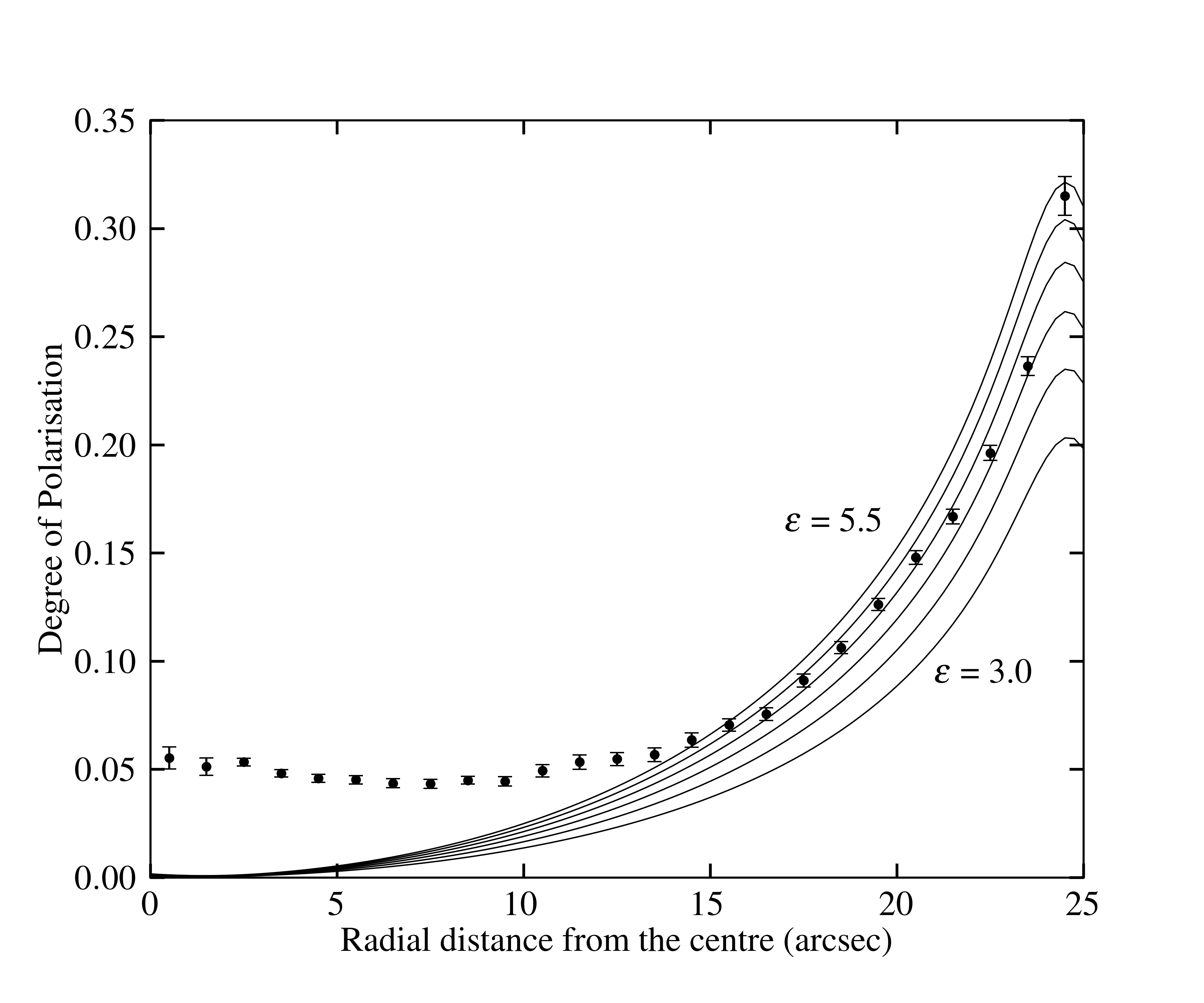}
    \captionsetup{font=footnotesize}
  \caption{
	   Variation of the observed degree of polarization with respect to the radial distance from the centre at 1297.67\,MHz as
	   observed on 06 September 2015 (solid dots). Solid lines show the theoretical curves convolved with Gaussian 
	   corresponding to the GMRT synthesized beam plotted for $\varepsilon$ = 3.0, 3.5, 4.0, 4.5, 5.0 and 5.5
 	   }
\label{fig:Fig8}
\end{figure}

\section{Discussions}
In Figure 1 are given observed values of the $\textit{T}_\textup{b}$ of Venus measured with the VLA by
\citet{Butler2001} and that from the GMRT observations. Combining the results 
of these two radio interferometers provide observations of Venus over a wide 
range of wavelength, from 1.3\,cm to 128\,cm. The timing of the presented 10 hour 
GMRT observations was optimised to be close to the inferior conjunction of the 
planet, which enabled an improvement in SNR by factor of five and an increase in 
the number of resolution elements across the disc of the planet by a factor $>$4, 
when compared to the previous observations. The $\textit{T}_\textup{b}$ of $<$321\,K obtained in the 
current GMRT campaign at 128\,cm is considerably lower than that reported by 
\citet{Mohan2017}, and that at 23\,cm is consistent with the earlier 
measurements by \citet{Butler2001}.  These analyses indicate that the thermal 
emission from the Venus surface and subsurface continues to decrease in the dcm 
and m wavelength domain of the electromagnetic spectrum. Several attempts have 
been made to explain the observed decreasing trend of the $\textit{T}_\textup{b}$ of Venus 
\citep{Barrett1964, Muhleman1979}. \citet{Butler2001} 
considered atmospheric opacity as well the surface temperature of Venus for 
explaining observed values of $\textit{T}_\textup{b}$ at cm wavelengths and got good agreement till 
the wavelength upto $\sim$6\,cm. However, they were not able to explain the 
observations at dcm and m wavelengths. \citet{Warnock1972} proposed a 
two-layer model, i.e. a layer of soil overlaying a regolith having a different 
dielectric values. They also noted that the observed decrease of $\textit{T}_\textup{b}$ at these 
wavelengths is unlikely to arise from a lower value of the physical temperature 
of the subsurface in view of thick overlying hot atmosphere. The constant 
surface temperature of $\sim$735\,K resulting from the Greenhouse effect is quite 
likely to heat the sub-surface of Venus over the millions of years. A radiative 
transfer (RT) analysis may be fruitful for explaining the observed decreasing 
values of $\textit{T}_\textup{b}$ at decimetre and metre wavelengths, considering the likely 
variation in the dielectric properties of the Venusian subsurface.

The mean value of DOP derived near the limb of Venus at 1297.67\,MHz (Fig. 3) is 
about 32\% compared to the value of $\sim$22\% at 607.67\,MHz (Fig. 2) owing to a higher 
angle of incidence due to the better angular resolution at 1298.76\,MHz. A DOP of 
3-5\% is observed close to the centre of the disk. Ideally, the DOP near the 
centre of the map is expected to be zero due to the rotational symmetry of the 
polarised radiation. However, the intensity of polarised emission, is given by  
P = $\sqrt{Q^2+U^2}$. This quantity can never be negative and hence there is a non-zero 
mean positive bias to the noise \citep{Chapman1986}. The noise on the polarised 
intensity near the centre is of the order of the background noise which is 
prevalent throughout the map.  This noise is $\sim$0.1 mJy for 607.67\,MHz and $\sim$0.15 
mJy for 1297.67\,MHz and leads to the observed small values of DOP close to the 
centre of the disc. In Figs. 4 and 5, the observed DOP begins to match with the 
theoretical values beyond $\sim$10 arcsec for July 25, 607.67\,MHz and for September 
06 at 1297.67\,MHz observations when the signal begins to dominate the noise.

The value of the Venus surface dielectric constant is estimated from the DOP  
maps of Venus generated from the GMRT dataset and the theoretical calculations 
to be, $\sim$4.5 both at 1297.67\,MHz and 607.67\,MHz. These are lowest frequency 
polarimetric observations of Venus, and can hence probe the subsurface deeper 
that has been possible before. Earlier estimates of dielectric constant have 
yielded varied from 2.5 to 5.0 \citep{Clark1965, Muhleman1979, Pettengill1988}.
Our observations are consistent with the radar 
observations of Venus by several workers including the Magellan observations 
determined the values of the dielectric constant to lie between 4 and 4.5 for 
the surface of Venus \citep{Pettengill1992}. The dielectric constant derived 
from the DOP observations depends on: 1) the relative surface roughness with 
respect to the probing frequencies \citep{Golden1979} and 2) dielectric constant 
variation within the penetration depth due to the increase in the bulk density 
\citep{Ulaby1990}. The effect of frequency on dielectric constant was ruled out by 
\citet{Campbell1969}, based on their measurements of the dielectric constant 
of planetary rocks including the basalt type over a wide range of frequency from 
450\,MHz to 35\,GHz. Owing to the deeper penetration depth (nearly ten times the 
wavelength) in a planetary dry surface at dcm and m wavelengths, a variation of 
the regolith bulk density of Venus is a valid possibility. As the real part of 
dielectric constant of the dry planetary regolith are related to only the bulk 
density, it can vary from $\sim$2.0 to $\sim$8.0 \citep{Ulaby1990} with the variation of 
bulk density from very loose (regolith) condition to highly packed condition 
close to rock material. Due to the presence of greenhouse gases for millions of 
years, any drastic variation in the temperature of Venus subsurface regolith up 
to tens of metres is very unlikely. Also at these wavelengths, the surface could 
be relatively smoother and the effect of surface roughness can be neglected, 
leading to the estimated value of dielectric very close to the real value. As 
the dielectric constant plays a major role in the observed $\textit{T}_\textup{b}$ at various 
frequencies, the current results form an essential input for the radiative 
transfer based models for exploring the role of subsurface properties 
(temperature and dielectric properties through density and mineral content) on 
thermal emissions and the decrease in $\textit{T}_\textup{b}$ with wavelength
at cm/dcm wavelengths.

\section{CONCLUSION}

From the observations of thermal emission from Venus surface made with the GMRT 
at 1297.67\,MHz (23\,cm), 607.67\,MHz (49\,cm) and 233.67\,MHz (1.28 m), the disc 
averaged brightness temperature, $\textit{T}_\textup{b}$, values were derived as 622$\pm$43\,K, 554$\pm$38\,K
and upper limit of 321\,K, respectively at the above frequencies. Considering 
also measurements made by previous workers using the VLA, it is found that $\textit{T}_\textup{b}$ is 
$\sim$505\,K at 1.3\,cm, peaks at $\sim$680\,K at a wavelength of 6\,cm and then decreases 
continuously with increasing wavelengths to $<$321\,K at 1.28 m. A satisfactory 
model to explain the observations should consider the likely penetration depth 
of radiation at decimeter and centimeter wavelengths in the dielectric medium of 
Venus. We also present the polarimetric observations of Venus at the lowest 
frequency set, which allows us to probe deeper into the subsurface. Using 
Stokes parameters derived from these observations, the DOP maps of Venus were 
generated at 1297.67\,MHz and 607.67\,MHz. A combined analysis of observed DOP and 
theoretically derived DOP, the dielectric constant of the surface of Venus was 
found to be $\sim$4.5 at both 1297.67\,MHz and 607.67\,MHz. This is consistent with 
those derived from the orbiter-based radar observations of Venus which lie in 
the range 4 - 4.5 for the surface of Venus. 

\section*{Acknowledgements}

Authors thank the Reviewer Dr Peter G. Ford, Assistant Editor and Scientific 
Editor for several valuable suggestions for improvements on this paper. Authors 
thank Dr. Subhasish Roy and Dr. Niruj Mohan Ramanujam of NCRA and Dr. Nizy 
Mathew of SPL, VSSC for their valuable comments and wholehearted support 
throughout the work. Authors thank the staff of the GMRT who made these 
observations possible. The GMRT is run by the National Centre for Radio 
Astrophysics of the Tata Institute of Fundamental Research. Mr. Nithin Mohan is 
supported by an ISRO Research Fellowship.

\bibliographystyle{mnras} 

%\bibliography{refer_final.bib}

\label{lastpage}

\end{document}